\documentclass[aps,pra,reprint,groupedaddress]{revtex4-1} 

\usepackage{amssymb}
\usepackage{amsmath,amsthm}
\usepackage[shortlabels]{enumitem}
\usepackage{hyperref}
\usepackage{uri} 
\usepackage{color}

\theoremstyle{plain}

\theoremstyle{definition}

\DeclareMathOperator{\trace}{tr}

\newcommand \bra[1] {\langle {#1} |}
\newcommand \ket[1] {| {#1} \rangle}

\newcommand \pairing[2] {({#1} | {#2})}

\newcommand\vrepDmat {\mathcal{A}}

\newcommand\Fcs {F_{\mathrm{cs}}}
\newcommand\FHxc {F_{\mathrm{Hxc}}}
\newcommand\Fxc {F_{\mathrm{xc}}}
\newcommand\FH {F_{\mathrm{H}}}
\newcommand\FSC {F_{\mathrm{SC}}}
\newcommand\Ts {T_s}
\newcommand\FmodHxc {\widetilde{F}_{\mathrm{Hxc}}}

\newcommand\Gcs {G_{\mathrm{cs}}}
\newcommand\GHxc {G_{\mathrm{Hxc}}}
\newcommand\GH {G_{\mathrm{H}}}
\newcommand\Gx {G_{\mathrm{x}}}
\newcommand\Gxc {G_{\mathrm{xc}}}
\newcommand\GmodHxc {\widetilde{G}_{\mathrm{Hxc}}}

\newcommand \Eloc {E_{\mu}}
\newcommand \NVRepGap {\Delta}
\newcommand{\vext}{v_\mathrm{ext}}
\newcommand{\vL}{v_\mathrm{L}}

\newcommand{\rmd}{\mathrm{d}}
\newcommand{\rmi}{\mathrm{i}}
\newcommand{\rr}{\mathbf{r}}

\begin{document}

\title{Fully self-consistent optimized effective potentials from a convex minimization problem}

\author{Erik I. Tellgren}
\email{erik.tellgren@kjemi.uio.no}
\affiliation{
	Hylleraas Centre for Quantum Molecular Sciences, Department of Chemistry, University of Oslo, P.O.~Box 1033 Blindern, N-0315 Oslo, Norway}

\author{Andre Laestadius}
\affiliation{
	Hylleraas Centre for Quantum Molecular Sciences, Department of Chemistry, University of Oslo, P.O.~Box 1033 Blindern, N-0315 Oslo, Norway}

\author{Markus Penz}
\affiliation{Basic Research Community for Physics, Leipzig, Germany}

\date{\today}

\begin{abstract}
The optimized effective potential method is formulated as a convex minimization problem. This formulation does not require assumptions about $v$-representability nor functional differentiability. The formulation provides a natural framework for fully self-consistent calculations, where both a Kohn--Sham system with a non-local potential and an additional system with a local potential are jointly optimized. The formulation is also well suited for extensions to other flavors of density-functional theory, e.g., current-density functional theory, where additional potentials besides the electrostatic potential are added.
\end{abstract}

\maketitle

\section{Introduction}

Most density-functional approximations can be classified on the rungs of ``Jacob's ladder'' based on the ingredients they use~\cite{PERDEW_AIPCP577_1}. The local density approximation (LDA) and the generalized gradient approximations (GGA) make up the two lowest rungs and are proper density functionals. At higher rungs more pragmatic functionals are found that depend on the Kohn--Sham (KS) orbitals~\cite{KUMMEL_RMP80_3}. Although this can be motivated using the Hohenberg--Kohn theorem, it has the disadvantage that the exchange-correlation (xc) contribution to the KS Hamiltonian is in practice a nonlocal operator, rather than the local potential that appears in the conventional KS equations. The nonlocality of the xc contribution affects the KS eigenvalue spectrum and in particular the HOMO-LUMO or band gap. The question then arises, how to compute local potentials and the corresponding KS spectrum for orbital-dependent xc functionals. The optimized effective potential (OEP) method aims at being a practically feasible solution to this problem. It has been formulated in several different ways by different authors, at various levels of generality and approximation. Sharp and Horton~\cite{SHARP_PR90_317}, and later Talman and Shadwick~\cite{TALMAN_PRA14_36}, sought to find a local approximation to the nonlocal Hartree--Fock exchange potential. With the emergence of functionals that go beyond the inclusion of exact exchange by including orbital dependence inspired by the Random Phase Approximation or Green's function methods, it has also become more interesting to develop the OEP method to a general method not specific to exact exchange~\cite{CASIDA_PRA51_2005,YANG_PRL89_143002,VOORA_PRA99_012518}.

In this work, we revisit the OEP method to provide a novel formulation that is fully variational, involves a simple minimization principle, and is suitable for fully self-consistent calculations (expressed in Eq.~\eqref{eqOEPMinPrincB}). That is, we construct a \emph{bivariational} minimization principle, which involves joint optimization of both, a system with a non-local potential and an auxiliary system with a local potential.

\section{Kohn--Sham theory}
\label{sec:KStheory}

Consider the standard quantum-chemical Hamiltonian
\begin{equation}\label{eq:def-H-lambda}
    \hat H^{\lambda}(v) = \hat T + \hat V + \lambda \hat W=-\frac{1}{2} \sum_j \nabla_j^2 + \sum_j v(\rr_j) + \sum_{j<l} \frac{\lambda}{r_{jl}},
\end{equation}
with $r_{jl}=|\rr_j-\rr_l|$, 
where $\lambda$ interpolates between a noninteracting system ($\lambda=0$) and one with full electron--electron repulsion ($\lambda=1$). 
The Levy--Lieb constrained-search functional is defined as~\cite{LEVY_PNAS76_6062,LIEB_IJQC24_243}
\begin{equation}
    \label{eq:def-Fcs}
    \Fcs^{\lambda}(\rho) = \inf_{\Gamma \mapsto \rho} \trace{\Gamma \hat H^{\lambda}(0)},
\end{equation}
where the optimization may be carried out either over only pure states $\Gamma = \ket{\Psi} \bra{\Psi}$ or over all mixed states. In the latter case, the functional is convex since the mapping $\Gamma\mapsto \rho$ is linear.
The well-known Rayleigh--Ritz variational principle,
\begin{equation}
    E^{\lambda}(v) = \inf_{\Gamma} \trace{\Gamma \hat H^{\lambda}(v)} = \inf_{\Psi} \bra{\Psi} \hat H^{\lambda}(v) \ket{\Psi},
\end{equation}
can now be rewritten into the Hohenberg--Kohn variational principle~\cite{HOHENBERG_PR136_864},
\begin{equation}
    E^{\lambda}(v) = \inf_{\rho} \big( \pairing{v}{\rho} + \Fcs^{\lambda}(\rho) \big),
\end{equation}
where compact notation is used for the pairing $\pairing{v}{\rho} = \int v(\rr) \rho(\rr) \, \rmd\rr$.
Somewhat less well-known is the dual Lieb variational principle~\cite{LIEB_IJQC24_243},
\begin{equation}
   \label{eqLiebVarPrinc}
    \Fcs^{\lambda}(\rho) = \sup_{v} \big(  E^{\lambda}(v) - \pairing{v}{\rho} \big).
\end{equation}

Let us now take as point of departure an interacting ($\lambda=1$) system subject to the external potential $\vext$. Let $F^1(\rho)$ denote either the exact, interacting functional $\Fcs^1(\rho)$ or any approximation to it. Likewise, we define a noninteracting kinetic-energy functional
\begin{equation}\label{eq:Ts-def}
    \Ts(\rho) = \inf_{\substack{\Gamma \in \mathcal{S}_N \\ \Gamma \mapsto \rho}} \trace{\Gamma \hat T},
\end{equation}
where the domain $\mathcal{S}_N$ either contains all valid $N$-body mixed states or only those that correspond to Slater determinants. The former choice corresponds to allowing fractional occupation of KS orbitals and implies $\Ts(\rho) = \Fcs^{0}(\rho)$ which guarantees that this functional is convex again, while the latter corresponds to forcing integer occupation numbers and implies $\Ts(\rho) \geq \Fcs^{0}(\rho)$.

Let $\vext$ be a fixed external potential. For an approximate functional $F^1(\rho)$, we define the ground-state energy in the same way as for the exact case above, i.e.,
\begin{equation}
    E^1(\vext) = \inf_{\rho} \big(
    \pairing{\vext}{\rho} + F^1(\rho) \big).
\end{equation}
For the exact functional and sufficiently accurate approximations, this is a convex optimization problem. Now we introduce a KS system described by a one-particle reduced density matrix, 
\begin{equation}
    D(\rr,\rr') = N \int \Gamma(\mathbf r,\mathbf r_2 \dots, \mathbf r_N; \mathbf r',\mathbf r_2\dots,\mathbf r_N) \rmd \rr_2 \cdots \rr_N ,
\end{equation}
and use this system to parametrize the density,
\begin{equation}
    E^1(\vext) = \inf_{D \in \mathcal{S}} \big( \pairing{\vext}{\rho_D} + F^1(\rho_D) \big).
\end{equation}
Here, $\mathcal{S}$ could be either all valid one-particle reduced density matrices or only the idempotent ones corresponding to Slater determinants. In the former case, $\mathcal{S}$ is convex and the reparametrization is linear, so the convexity of the original optimization problem is preserved. However, the main motivation for introducing a KS system is not to obtain the above form, but to circumvent the difficulty of modeling the kinetic-energy contribution $\Ts(\rho)$ to $F^{1}(\rho)$. It is easier to model the Hartree-exchange-correlation (Hxc) functional, defined by
\begin{equation}
    \FHxc(\rho) = F^1(\rho) - \Ts(\rho)
\end{equation}
and typically further decomposed into a Hartree $\FH(\rho)$ and an xc contribution $\Fxc(\rho) = \FHxc(\rho) - \FH(\rho)$, with
\begin{equation}
    \FH(\rho) = \frac{1}{2} \int \frac{\rho(\rr_1) \, \rho(\rr_2)}{r_{12}} d\rr_1 d\rr_2.
\end{equation}
Because $\Ts(\rho_D) \leq \inf_{\Gamma\mapsto D}\trace{\Gamma\hat T} = \trace{D T}$ with $T = \tfrac{1}{2} \nabla\cdot\nabla'$ and since this inequality can always be saturated by a certain $D \mapsto \rho_D$, it now follows that
\begin{equation}
 \begin{split}
    E^1(\vext) & = \inf_{D \in \mathcal{S}} \big( \pairing{\vext}{\rho_D} + \Ts(\rho_D) + \FHxc(\rho_D) \big)
              \\
    & = \inf_{D \in \mathcal{S}} \big( \pairing{\vext}{\rho_D} + \trace{D T} + \FHxc(\rho_D) \big).
    \label{eqEdftReparam}
 \end{split}
\end{equation}
Writing a tilde on the functional, $\FmodHxc$, this indicates that next to the exact functional we also allow for any approximation to it.

The KS orbitals defining the minimizer $D^{\star}(\rr,\rr') = \sum_j n_j \phi_j(\rr) \phi_j(\rr')^*$ solve the KS equations
\begin{equation}
    \Big( -\frac{1}{2} \nabla^2 + v_{\mathrm{s}}(\rr) \Big) \phi_j(\rr) = \kappa_j \phi_j(\rr)
\end{equation}
with the local KS potential formally given by
\begin{equation}
    v_{\mathrm{s}}(\rr) = \vext(\rr) + \frac{\delta \FHxc(\rho)}{\delta \rho(\rr)}
\end{equation}
at the ground-state density.
Strictly speaking, though many approximate functionals allow functional derivatives to be defined, the exact functional is not differentiable~\cite{LAMMERT_IJQC} (but see Refs.~\onlinecite{kvaal2014differentiable,LAESTADIUS_JCP149_164103}). 
For any convex functional $F^1(\rho)$ the notion of a subgradient may be used instead, in order to achieve a rigorous treatment of potentials. We omit this technical point as it does not substantially impact the discussion of OEP.

Finally, we note that although the minimization domain $\mathcal{S}$ can be chosen as a convex set, the second line of Eq.~\eqref{eqEdftReparam} is no longer a convex optimization problem. The reason is that the functional $\Ts(\rho_D)$, which is convex in $D$ if $\mathcal{S}$ is convex, has been replaced by a term that is linear in $D$. The only nonlinear term is then a difference of two convex functions, $\FHxc(\rho_D) = F^1(\rho_D) - \Ts(\rho_D)$. Hence, convexity will usually be lost when the Hohenberg--Kohn minimization principle is reparametrized using a KS density matrix, even if the exact density functional is used. 
Notably, the above loss of convexity does not necessarily occur in the parallel formulation of density-matrix functional theory defined later in Eq.~\eqref{eqDMFTvarHK}, as there the kinetic-energy functional is manifestly linear in $D$. In the exact case, the nonconvexity in the DFT setting can be bounded by introducing the strictly correlated functional~\cite{SEIDL_PRA59_51},
\begin{equation}
    \FSC(\rho) = \inf_{\Gamma \mapsto \rho} \trace{\Gamma \hat W}.
\end{equation}
Now, since $\Fcs^1(\rho) \geq \Fcs^0(\rho) + \FSC(\rho)$ and $\FHxc(\rho) \geq \FSC(\rho)$, we obtain
\begin{equation}
    E^1(\vext) \geq \inf_{D} \big( \pairing{\vext}{\rho_D} + \trace{D T} + \FSC(\rho_D) \big).
\end{equation}
The right-hand side is again a convex optimization problem, since $\FSC(\rho_D)$ is a convex functional.

\section{Orbital-dependent functionals}

We shall use the one-particle reduced density matrix,
\begin{equation}
    D(\rr,\rr') = \sum_j n_j \phi_j(\rr) \phi_j(\rr')^*,
\end{equation}
as a convenient collective representation of a set of orthonormal orbitals $\phi_j$ and occupation numbers $n_j \in [0,2]$. Note that we consider a spin-restricted formalism---the generalization to an explicit consideration of spin is straightforward. In what follows, we also leave it unspecified whether the occupation numbers $n_j$ are allowed to be fractional or must be integers (0 or 2). Any orbital-dependent functional $G^{\lambda}(\{(n_j,\phi_j)\})$ can now be reinterpreted as a density-matrix functional $G^{\lambda}(D)$, provided that it is invariant under rotations within degenerate subspaces of orbitals sharing the same occupation number.

Next we introduce the exact density-matrix functional through the constrained-search expression
\begin{equation}
    \Gcs^{\lambda}(D) = \inf_{\substack{\Gamma \in \mathcal{S}_N \\ \Gamma \mapsto D}} \trace{\Gamma \hat H^{\lambda}(0)}.
\end{equation}
Because the kinetic-energy term is directly determined by $D$,
\begin{equation}\label{eq:G0cs}
\begin{aligned}
    \Gcs^{0}(D) &= \inf_{\substack{\Gamma \in \mathcal{S}_N \\ \Gamma \mapsto D}} \trace{\Gamma \hat T} = \trace{D T} \\
    &= \frac{1}{2} \int \nabla\cdot\nabla' D(\rr,\rr') \Big|_{\rr'=\rr} \rmd\rr,
\end{aligned}
\end{equation}
one immediately has
\begin{equation}\label{eq:GHxc-from-W}
    \GHxc(D) = \Gcs^{1}(D) - \Gcs^{0}(D) = \inf_{\substack{\Gamma \in \mathcal{S}_N \\ \Gamma \mapsto D}} \trace{\Gamma \hat W}.
\end{equation}
This density-matrix functional resembles the strictly correlated density functional discussed in Sec.~\ref{sec:KStheory}. Moreover, the Hartree and exact exchange contributions to $\GHxc(D)$ are easy to separate out as well,
\begin{equation}
    \GH(D) = \FH(\rho_D) = \frac{1}{2} \int \frac{\rho_D(\rr_1) \, \rho_D(\rr_2)}{r_{12}} \rmd\rr_1 \rmd\rr_2
\end{equation}
and
\begin{equation}
    \Gx(D) = -\frac{1}{4} \int \frac{|D(\rr_1,\rr_2)|^2}{r_{12}} \rmd\rr_1 \rmd\rr_2.
\end{equation}

Next, let $\GmodHxc(D)$ be either the exact, interacting Hxc functional or any approximation to it. For example, it could be a meta-GGA model, a model that only includes Hartree and exact exchange, or it could contain a correlation model as well. The total functional is taken to be 
\begin{equation}
    \label{eq:G-functional}
    G^{\lambda}(D) = \Gcs^0(D) + \lambda \GmodHxc(D).    
\end{equation}
The ground-state energy is then
\begin{equation}
   \label{eqDMFTvarHK}
   E^{\lambda}(v) = \inf_{D \in \mathcal{S}} \big( \pairing{v}{\rho_D} + G^{\lambda}(D) \big).
\end{equation} 
In order for the exact functional to reproduce the exact ground-state energy, the minimization domain $\mathcal{S}$ needs to contain precisely the valid density matrices $D$, including those that have fractional occupation numbers. However, as an approximate model, one has the option of only minimizing over density matrices with integer occupation. We thus continue leaving the exact minimization domain unspecified. The minimizing $D^{\star}$ defines a kind of KS system that is subject to a nonlocal Hxc potential that we may indicate as the functional derivative
\begin{equation}\label{eq:GHxc-diff}
    v_{\mathrm{Hxc}}(\rr',\rr) = \left. \frac{\delta \GmodHxc(D)}{\delta D(\rr,\rr')} \right|_{D=D^{\star}}.
\end{equation}
The orbitals now satisfy the nonlocal KS-like equation
\begin{equation}
    -\frac{1}{2} \nabla^2 \phi_j(\rr) + \int v_{\mathrm{s}}(\rr,\rr') \, \phi_j(\rr') \, d\rr' = \varepsilon_j \phi_j(\rr),
\end{equation}
with $v_{\mathrm{s}}(\rr,\rr') = \delta(\rr-\rr') \vext(\rr) + v_{\mathrm{Hxc}}(\rr,\rr')$. In particular, the spectrum of orbital energies is not the same as that obtained from a local KS potential.

Formally, any density-matrix functional can be turned into a density functional by identifying the functional
\begin{equation}
    F^{\lambda}(\rho) = \inf_{D \mapsto \rho} G^{\lambda}(D)
\end{equation}
in the nested minimization
\begin{equation}
   \label{eqDFTVarHKfromDMFT}
   E^{\lambda}(v) = \inf_{\rho} \Big( \pairing{v}{\rho} + \inf_{D \mapsto \rho} G^{\lambda}(D) \Big).
\end{equation}
It is, however, not practical to directly implement a constrained search over $D \mapsto \rho$. Yang and Wu~\cite{YANG_PRL89_143002} suggested to obtain the density functional from the Lieb maximization principle
\begin{equation}
    F^{\lambda}(\rho) = \sup_v \big( E^{\lambda}(v) - \pairing{v}{\rho} \big).
\end{equation}
However, when inserted into the Hohenberg--Kohn-like variation principle in Eqs.~\eqref{eqDMFTvarHK} or  \eqref{eqDFTVarHKfromDMFT}, a maximin optimization problem results. Therefore, this procedure is mainly suitable for post hoc calculations, after the minimizing $D$ and $\rho_D$ have already been obtained. In order to achieve a form more suitable for fully self-consistent calculations, a pure minimization problem is desirable.

\section{The Optimized Effective Potential method}

The OEP method is often presented by using response functions and the chain rule for functional differentiation. Here, however, following Yang and Wu~\cite{YANG_PRL89_143002}, we understand the OEP method as the problem of finding the \emph{local} potential $\vL$ such that the lowest orbitals $\zeta_j$ arising from the single-particle equations
\begin{equation}
    \big( -\tfrac{1}{2} \nabla^2 + \vL(\rr) \big) \zeta_j(\rr) = \kappa_k \zeta_j(\rr)
\end{equation}
yield the lowest energy of an orbital-dependent model such as an exact-exchange energy functional or even a orbital-dependent correlation functional~\cite{KUMMEL_RMP80_3}. 
This means that additionally to the fully interacting system with $\vext(\rr)$ and the KS system with a \emph{non}-local $v_s(\rr,\rr')$, a third system is introduced with a \emph{local} potential $\vL(\rr)$ that when optimized reproduces the same density.
Letting $\mathcal{P}(v)$ denote the projector on the lowest $N$ orbitals (or $N/2$ doubly occupied orbitals), the problem can be written
\begin{equation}\label{eq:OEP-var-principle}
 \begin{split}
    E_{\mathrm{OEP}}^1(\vext) & = \min_v \Big( \pairing{\vext}{\rho_{\mathcal{P}(v)}} + G^1(\mathcal{P}(v)) \Big)
        \\
    & = \min_{P \in \vrepDmat} \Big( \pairing{\vext}{\rho_{P}} + G^1(P) \Big),
 \end{split}
\end{equation}
where $\vrepDmat$ is the set of noninteracting $v$-representable density matrices and $P\in\vrepDmat$ represents the chosen orbitals. Relatedly, the above expression features a minimum rather than an infimum, since standard OEP treatments assume $v$-representability anyway.
Clearly, $E_{\mathrm{OEP}}^1(\vext) \geq E^1(\vext)$, since the domain of reduced density matrices $\vrepDmat \subset \mathcal{S}$ is effectively restricted by the use of $v$ as a variational parameter. We leave the issue of how to handle degeneracies at the Fermi level of the KS system unspecified, as it does not matter for the following discussion.

The fact that the orbital-dependent functional is now a composition $G^1 \circ \mathcal{P}$ means that one could also rely on the chain rule to differentiate the total energy with respect to $v$. This has the disadvantage of introducing a singular response function into even the gradient expression~\cite{HIRATA_JCP115_1635,DELLASALA_JCP115_5718}. Moreover, the mathematical properties of the composition $G^1 \circ \mathcal{P}$ are not obvious.
The aim will thus be to come up with an alternative variational principle to Eq.~\eqref{eq:OEP-var-principle} with better properties that still yields the local potential.

\section{Minimization principle for self-consistent OEP}

Trivially, from the definition of $E^{\lambda}(v)$, we have $E^{\lambda}(v) \leq \pairing{v}{\rho_D} + G^{\lambda}(D)$ for all $v$ and $D$ and we follow Lammert~\cite{LammertBivariate} in defining the nonnegative {\it excess energy},
\begin{equation}
    Q^{\lambda}(v,D) = \pairing{v}{\rho_D} + G^{\lambda}(D) - E^{\lambda}(v) \geq 0.
\end{equation}
The excess energy vanishes whenever $D$ is the correct ground-state density matrix for the potential $v$. For the noninteracting system, we have the particularly simple form
\begin{equation}
    Q^{0}(v,D) = \pairing{v}{\rho_D} + \trace{D T} - E^{0}(v) \geq 0.
\end{equation}

We now fix a parameter $\mu > 0$ and define the modified energy
\begin{equation}
    \label{eqOEPMinPrincA}
    \Eloc^{1}(\vext) = \inf_{D,\vL} \big( \pairing{\vext}{\rho_D} + G^{1}(D) + \mu Q^0(\vL,D) \big),
\end{equation}
where the subscripts `ext' and `L' distinguish the potentials for the full interacting and the local noninteracting systems. The modified energy never underestimates the unmodified energy, $E^1(\vext) \leq \Eloc^1(\vext)$. 

Some further insight into the above modified energy is provided by introducing the \emph{v-representability gap} of a density matrix,
\begin{equation}
    \NVRepGap^{\lambda}(D) = \inf_v Q^{\lambda}(v,D) \geq 0,
\end{equation}
which vanishes when the density matrix is $v$-representable. Then we can write
\begin{equation}
    \label{eqOEPMinVRepInterp}
    \Eloc^{1}(\vext) = \inf_{D} \big( \pairing{\vext}{\rho_D} + G^{1}(D) + \mu \NVRepGap^{0}(D) \big).
\end{equation}
Hence, the modified energy is a compromise between the standard energy and the energy penalty $\NVRepGap^0(D)$ for failures of noninteracting $v$-representability of $D$.
The difference to the unmodified energy can also be written in the more symmetric form
\begin{equation}
    \Eloc^1(\vext) - E^1(\vext) = \inf_{D,\vL} \big( Q^1(\vext,D)  + \mu Q^0(\vL,D) \big).
\end{equation}

By writing out the modified energy more explicitly,
\begin{equation}
 \label{eqOEPMinPrincB}
 \begin{split}
    \Eloc^1(\vext) & = \inf_{D,\vL} \big( \pairing{\vext+\mu \vL}{\rho_D} + (1+\mu) \trace{D T}
          \\
    & \quad \quad \quad \quad + \GmodHxc(D) - \mu E^0(\vL) \big),
 \end{split}
\end{equation}
we arrive at the desired result of this work, a joint minimization principle for the density matrix and the local potential.
Its minimizer $D^\star$ is not quite a ground state for $\vext$ but yields the \emph{modified} energy $\Eloc^1(\vext) \geq E^1(\vext)$ instead, while we also get an optimal, effective \emph{local} potential $\vL^\star$ (depending on $\mu$) as the minimizing potential.
Variation of Eq.~\eqref{eqOEPMinPrincB} over $\vL$ at $D=D^\star$ yields the stationarity condition
\begin{equation}\label{eq:vL-stationarity}
    \rho_{D^\star} - \left.\frac{\delta E^0(\vL)}{\delta \vL}\right|_{\vL=\vL^\star}  = \rho_{D^\star} - \rho_{P^{\star}} = 0
\end{equation}
that shows that the densities $\rho_{D^\star} = \rho_{P^{\star}}$ agree.
Here $\rho_{P^{\star}}$ is the ground-state density associated with the noninteracting Hamiltonian $\hat H^0(\vL^\star)$ and the energy $E^0(\vL^\star)$.
The orbitals that define the minimizer $D^{\star}(\rr,\rr') = \sum_j n_j \phi_j(\rr) \phi_j(\rr')^*$ satisfy the modified KS equation
\begin{equation}\label{eq:KS-Dstar}
 \begin{split}
    \Big( -\frac{1+\mu}{2} \nabla^2 &+ \vext(\rr) + \mu \vL(\rr) \Big) \phi_j(\rr) \\
    &+ \int v_{\mathrm{Hxc}}(\rr,\rr') \phi_j(\rr') \rmd\rr'
    = \varepsilon_j \phi_j(\rr),
 \end{split}
\end{equation}
at the minimizer $\vL = \vL^{\star}$ that is the result of a variation of Eq.~\eqref{eqOEPMinPrincB} over $\phi_j^*$.
Here the kinetic-energy operator is weighted by an unconventional factor $1+\mu$ and a nonlocal potential remains. However, by the usual Hohenberg--Kohn argument and motivated by Eq.~\eqref{eq:vL-stationarity} there is now a second, noninteracting system described by a density matrix $P^{\star}(\rr,\rr') = \sum_j m_j \zeta_j(\rr) \zeta_j(\rr')^*$ which realizes the minimum energy $E^0(\vL^\star) = \trace{P^{\star} T} + \pairing{\vL^\star}{\rho_{P^{\star}}}$ and shares the same density $\rho_{P^{\star}} =\rho_{D^{\star}}$. The orbitals of this system satisfy the local KS equation
\begin{equation}\label{eq:KS-Pstar}
    \Big( -\frac{1}{2} \nabla^2 + \vL(\rr) \Big) \zeta_j(\rr) = \kappa_j \zeta_j(\rr),
\end{equation}
again at the minimizer $\vL = \vL^{\star}$.
In practice, Eqs.~\eqref{eq:KS-Dstar} and \eqref{eq:KS-Pstar} have to be solved self-consistently, since we cannot compute the optimal $D^{\star}$ and $\vL^{\star}$ directly. Instead, in a simple self-consistent procedure, Eqs.~\eqref{eq:KS-Dstar} lets us obtain new orbitals and an update to $D$. Moreover, Eq.~\eqref{eq:KS-Pstar} allows us obtain $\rho_P$, which is needed e.g. to compute the gradient with respect parameters in $\vL$ (see Sec.~\ref{secFinBasForm}).

In the event that the density matrix $D^{\star}$ is noninteracting $v$-representable, the second system coincides with the first in the sense that $D^{\star}=P^{\star}$. It also follows that all occupation numbers, sorted in descending order, must agree, $n_j = m_j$. Moreover, all orbitals with unique occupation numbers coincide, $\phi_j = \zeta_j$, if there is strict inequality $n_{j-1} < n_j < n_{j+1}$. Individual orbitals that belong to degenerate subspaces can, however, be different.
The same happens if $\mu\to\infty$, since then in the Hamiltonian of Eq.~\eqref{eq:KS-Dstar} only the terms from Eq.~\eqref{eq:KS-Pstar} survive.

Therefore, we expect that in the opposite limit $\mu \to 0^+$ the perturbation of the optimal density matrix $D^{\star}$ is minimal. If, at $\mu=0$, $\rho_{D^{\star}}$ is noninteracting $v$-representable, then the same density matrix is obtained also in the limit $\mu \to 0^+$. On the other hand, if there is a failure of noninteracting $v$-representability at $\mu=0$, the limit $\mu \to 0^+$ yields a minimally disturbed $D^{\star}$.

In general, $E^0(\vL)$ is a concave functional of $\vL$. As a consequence, the minimization over $\vL$ in Eq.~\eqref{eqOEPMinPrincB} has the virtue of defining a convex optimization problem {\it irrespective of the choice of model functional} $\GmodHxc(D)$. If, in addition, the model Hxc functional is convex in $D$, then even the joint minimization over $(D,\vL)$ is a \emph{fully convex} optimization problem.

Additionally, the orbitals $\zeta_j$ and potential $\vL$ have all the advantages of conventional OEP formulations. Unlike the nonlocal potential, the local $\vL$ can be viewed as an approximate KS potential. Moreover, the resulting eigenvalue spectrum has different properties than the spectrum arising from a nonlocal potential. For example, exact exchange differentiates between occupied and unoccupied orbitals in a way that a local potential cannot.

\section{Finite basis formulation}
\label{secFinBasForm}

To construct a practical scheme, a standard orbital basis set $\{\chi_{\gamma}(\rr)\}_{\gamma}$ can be introduced,
\begin{align}
    \phi_j(\rr) & = \sum_{\gamma} C_{\gamma j} \, \chi_{\gamma}(\rr),
        \\
    \zeta_j(\rr) & = \sum_{\gamma} C^{\mathrm{L}}_{\gamma j} \, \chi_{\gamma}(\rr).
\end{align}
Alongside the orbital basis set, following the approach by Yang and Wu~\cite{YANG_PRL89_143002}, a potential basis set $\{\omega_{\beta}\}_{\beta}$ is used to expand the local potential
\begin{equation}
    \vL(\rr) = v_{\mathrm{ref}}(\rr) + \sum_{\beta} V^{\mathrm{L}}_{\beta} \, \omega_{\beta}(\rr).
\end{equation}
A fixed reference potential $v_{\mathrm{ref}}$, such as the Fermi--Amaldi potential, may also be included to e.g.\ ensure the correct long-range decay properties. In practice, the variational parameters are then the coefficients $C_{\gamma j}$ and $V^{\mathrm{L}}_{\beta}$. The energy $E^0(\vL)$ can be obtained by diagonalizing the basis set representation of the local Hamiltonian $\hat H^0(\vL)$,
\begin{equation}
    \mathcal{H}_{\gamma \eta} = \bra{\chi_{\gamma}} -\tfrac{1}{2} \nabla^2 + \vL \ket{\chi_{\eta}}.
\end{equation}
The gradient of the modified energy expression with respect to the expansion coefficients for the local potential is given by
\begin{equation}
    \frac{\partial \Eloc^1}{\partial V^{\mathrm{L}}_{\beta}} = \mu \pairing{\omega_{\beta}}{\rho_D - \rho_P}.
\end{equation}
From second order perturbation theory, one also gets the exact Hessian with respect to $V_{\beta}$ as
\begin{equation}
    \frac{\partial^2 \Eloc^1}{\partial V^{\mathrm{L}}_{\beta} \partial V^{\mathrm{L}}_{\gamma}} = -\frac{\mu^2}{2} \sum_{j,a} \frac{\bra{\zeta_j} \omega_{\beta} \ket{\zeta_a} \bra{\zeta_a} \omega_{\gamma} \ket{\zeta_j}}{\kappa_a - \kappa_j} + \mathrm{c.c.},
\end{equation}
where $j$ ($a$) now runs over occupied (unoccupied) orbitals and it is assumed that no orbital is fractionally occupied. The Hessian is negative semidefinite and becomes ill-defined when the noninteracting system described by $\hat H^{0}(\vL)$ has a ground-state degeneracy.

When the gradient vanishes, that is
\begin{equation}
    \frac{\partial \Eloc^1}{\partial V^{\mathrm{L}}_{\beta}} = \mu \pairing{\omega_{\beta}}{\rho_D - \rho_P} = 0
\end{equation}
for all $\beta$, the density $\rho_D$ from the orbital-dependent model and the density $\rho_P$ share the same projection in the space spanned by $\{\omega_{\beta}\}_\beta$. When this holds we say that {\it $\rho_D$ has a noninteracting $v$-representable projection}. As has been discussed extensively in the literature~\cite{KOLLMAR_JCP127_114104,HESSELMANN_JCP127_054102,HEATONBURGESS_JCP129_194102,JACOB_JCP135_244102}, the matching of the projections onto all $\omega_{\beta}$ does not mean that $\vL$ is visually close to its true basis set limit as a function of $\rr$. The real-space representation of the potential $\vL(\rr)$ may contain spurious oscillations that are ``invisible'' to the energy minimization due to negligible projection in the space spanned by orbital basis function products $\chi_{\gamma} \chi_{\eta}^*$.

\section{Specialization to hybrid functionals}

In the important case of hybrid density functionals, the orbital-dependent model takes the form
\begin{equation}
    \GmodHxc(D) = \FH(\rho_D) + \alpha \Gx(D) + \widetilde{F}_{\mathrm{GGA}}(\rho_D),
\end{equation}
where $\alpha$ is the fraction of exact exchange and the xc functional,
\begin{equation}
    \widetilde{F}_{\mathrm{GGA}}(\rho) = \int f_{\mathrm{GGA}}(\rho(\rr),\nabla\rho(\rr)) \, \rmd\rr,
\end{equation}
is a proper density functional of GGA type. Setting $\alpha=1$ and $\widetilde{F}_{\mathrm{GGA}}(\rho) = 0$ recovers Hartree--Fock theory as a special case.

Since the only truly orbital-dependent term is the exact exchange, the OEP problem now reduces to finding the local potential that optimally accounts for the nonlocal exact exchange. The total potential takes the form
\begin{equation}
 \begin{split}
    v_{\mathrm{Hxc}}(\rr,\rr') & = \frac{\delta \GmodHxc(D)}{\delta D} 
             \\
   & = \big(v_{\mathrm{H}}(\rr)+\widetilde{v}_{\mathrm{GGA}}(\rr)\big) \delta(\rr-\rr') + \alpha v_{\mathrm{x}}(\rr,\rr')
 \end{split}
\end{equation}
with the contributions
\begin{align}
    v_{\mathrm{H}}(\rr) & = \frac{\delta \FH(\rho)}{\delta \rho(\rr)} = \int \frac{\rho(\rr')}{|\rr-\rr'|} \rmd \rr',
            \\
    \widetilde{v}_{\mathrm{GGA}}(\rr) & = \frac{\delta \widetilde{F}_{\mathrm{GGA}}(\rho)}{\delta \rho(\rr)} = \frac{\partial f_{\mathrm{GGA}}}{\partial \rho(\rr)} - \nabla\cdot\frac{\partial f_{\mathrm{GGA}}}{\partial (\nabla\rho(\rr))},
            \\
    v_{\mathrm{x}}(\rr',\rr) & = \frac{\delta \Gx(D)}{\delta D(\rr,\rr')}.
\end{align}
As the first two potentials vary both with position and the density, we may alternatively write $v_{\mathrm{H}}(\rr;\rho)$ and $\widetilde{v}_{\mathrm{GGA}}(\rr;\rho)$, or even just $v_{\mathrm{H}}(\rho)$ and $\widetilde{v}_{\mathrm{GGA}}(\rho)$, to emphasize that they are density dependent.
In the reference potential, one can now consider including the Hartree and GGA potentials,
\begin{equation}
    v_{\mathrm{ref}}(\rr) = v_{\mathrm{H}}(\rr;\rho_D) + v_{\mathrm{GGA}}(\rr;\rho_D).
\end{equation}
The basis expansion is then completely targeted to the exchange term. This approach leads to the complication that the local potential $\vL = \vL(\rho_D)$ acquires an explicit dependence on $\rho_D$, which leads to an extra $\rho_D$-dependence in the energies
\begin{equation}
    \Eloc^1(\vext) = \inf_{D,V^{\mathrm{L}}} \big( \pairing{\vext}{\rho_D} + G^1(D) + \mu Q^0(\vL(\rho_D),D) \big)
\end{equation}
and
\begin{equation}
    Q^0(\vL(\rho_D), D) = \trace{D T} + \pairing{\vL(\rho_D)}{\rho_D} - E^0(\vL(\rho_D)).
\end{equation}
There will consequently be an extra contribution to the nonlocal KS operator and Hxc potential given by
\begin{equation}
    \mu \frac{\delta Q^0}{\delta \vL} \frac{\delta \vL(\rho_D)}{\delta D} = \mu (v_{\mathrm{H}}(\rho_D) - v_{\mathrm{H}}(\rho_{P}) + \ldots) .
\end{equation}

In the case of Hartree--Fock theory, the minimization over $D$ is at the outset restricted to the non-convex set of idempotent density matrices. However, it is known that the Hartree--Fock energy minimization can be relaxed to the convex domain of positive semidefinite density matrices~\cite{CANCES_IJQC79_82,CANCES_MMNA34_749}. When minimization is performed over all Hermitian $D \geq 0$ with the correct total occupation, the above minimization has a very special structure: It is a convex optimization problem w.r.t.\ $\vL$ and a quadratic optimization problem w.r.t.\ $D$. Unfortunately, the Hartree--Fock functional is \emph{not} convex in $D$, but the convex-quadratic structure is rich enough to guide the choice of algorithm and convergence analysis.

From the pragmatic perspective of self-consistent field iterations, the above formulation allows for different nesting of the optimizations over $D$ and $v_{\mathrm{L}}$: For every candidate $D$, one may calculate the optimal $v_{\mathrm{L}}$, e.g., by gradient or second-order optimization as indicated above. Alternatively, for every candidate $v_{\mathrm{L}}$, one may calculate the optimal $D$ through self-consistent field iterations based on only Eq.~\eqref{eq:KS-Dstar}. Finally, one may perform simultaneous optimization over the pair $(D,v_{\mathrm{L}})$ or construct a pragmatic hybrid scheme where one of the variables is updated less frequently than the other.

\section{Virial relations}

In this section we will derive virial relations akin to the well-known results in DFT due to \citet{LevyPerdewPhysRevA1985}. 
Such relations have the advantage of establishing a direct relation between the xc potentials and their respective energy functionals without depending on the differentiability of the functional as in Eq.~\eqref{eq:GHxc-diff}.
The virial theorem itself is a direct consequence of the Ehrenfest theorem for the expectation value of the operator $\sum_i \rr_i\cdot\nabla_i$ with respect to a ground state (or, more generally, eigenstate) $\Gamma$ of the interacting $\hat H^1$ from Eq.~\eqref{eq:def-H-lambda}. We thus have
\begin{equation}
    0 = \frac{\rmd}{\rmd t} \trace \left( \Gamma \sum_i (\rr_i\cdot\nabla_i)\right) = \rmi \sum_i \trace \left( \Gamma \left[ \hat H^1, \rr_i\cdot\nabla_i \right] \right)
\end{equation}
and from an evaluation of the commutator with the different parts of the standard quantum-chemical Hamiltonian the virial relation
\begin{equation}
  \label{eq:virialA}
    2\trace{\Gamma \hat T} = \sum_{i}\trace{ \left(\Gamma (\rr_i\cdot \nabla_i) ( \hat W + \hat V)\right) }.
\end{equation}
If the electron--electron interaction is modeled by the Coulomb repulsion, then we can use the identity $\sum_i(\rr_i\cdot \nabla_i)r_{jl}^{\alpha} = \alpha r_{jl}^{\alpha}$ and get
\begin{equation}
  \label{eq:virialB}
    2\trace{\Gamma\hat T } = - \trace{\Gamma\hat W}+ \sum_{i}\trace{\left(\Gamma (\rr_i\cdot \nabla_i) \hat V\right) }.
\end{equation}
Performing the infimum over all states with the restriction $\Gamma \mapsto D$ allows us to introduce the Hxc functional from Eq.~\eqref{eq:GHxc-from-W} as well as the density $\rho_D$ for the last trace, and we have
\begin{equation}\label{eq:VT2}
    2\trace D T = - \GHxc(D) + \int \rho_D(\rr) \,\rr\cdot\nabla v(\rr) \rmd\rr.
\end{equation}
The same result holds for the noninteracting case where the $\GHxc$ just drops.

Now, recall Eq.~\eqref{eqOEPMinPrincB}, which can be written
\begin{equation}
 \begin{split}
    \Eloc^1(\vext) & = \inf_{D,\vL} \big( 
    \trace{(\Gamma_D \hat H^1_\mu)}- \mu E^0(\vL) \big),
 \end{split}
\end{equation}
with the Hamiltonian
\begin{equation}
    \hat H^1_\mu = (1+\mu) \hat T + \sum_j \big[ \vext(\rr_j)+\mu \vL(\rr_j) \big] + \hat W .
\end{equation}
Let $D^\star$ be the minimizer of Eq.~\eqref{eqOEPMinPrincB} with the exact, interacting Hxc functional which makes it the ground 
state of $\hat H^1_\mu$. 
Then for the $\mu$-modified system, 
the virial theorem gives
\begin{align}\label{eq:VR-1st-part}
    &2(1+\mu) \trace{ D^\star T} \\
    &= - G_\mathrm{Hxc}(D^\star) + \int \rho_{D^\star}(\rr) \, \rr\cdot \nabla( \vext(\rr) +  \mu \vL^\star(\rr) )\rmd \rr. \nonumber
\end{align}
Here, $\vL^\star$ 
is the local potential from Eq.~\eqref{eq:KS-Pstar} such that
\begin{equation}
\rho_{P^\star} = \rho_{D^\star} 
= \rho.
\end{equation}
Note that in the limit $\mu\to \infty$ we furthermore achieve  $D^\star=P^\star$.

On the other hand, using the virial theorem for the system with potential $\vL^\star$ and ground state $P^\star$ we have
\begin{equation}\label{eq:VR-2nd-part}
    2 \trace{P^\star \hat T}
    = \int \rho(\rr) \,\rr\cdot \nabla \vL^\star(\rr) \rmd \rr.
\end{equation}
We define 
\begin{equation}\label{eq:TcD}
    T_{\mathrm{c}}(D^\star) = \trace{(D^\star - P^\star) T}
\end{equation}
where the part involving $P^\star$ gets fixed by choosing $D^\star$ since it comes from the corresponding KS system Eq.\eqref{eq:KS-Dstar}.
Now, combining Eqs.~\eqref{eq:VR-1st-part}, \eqref{eq:VR-2nd-part} and \eqref{eq:TcD} we obtain
\begin{align}
    \GHxc&(D^\star)  + 2(1+\mu) T_\mathrm{c}(D^\star) \nonumber \\
    &=  - \int \rho(\rr)\, \rr\cdot \nabla(  \vL^\star(\rr) -\vext(\rr) )\rmd \rr \\
    &=  \GH(\rho) 
    - \int \rho(\rr) \,\rr \cdot \nabla(\vL^\star(\rr) - v_\mathrm{H}(\rr)-\vext(\rr) )\rmd \rr. \nonumber
\end{align}
Set
$v_{\mathrm{L,xc}}^\star = \vL^\star - v_\mathrm{H}-\vext,$
such that we obtain the main result in this section,
\begin{equation}\label{eq:VR-loc-xc}
    \Gxc(D^\star) 
    + 2(1+\mu) T_\mathrm{c}(D^\star) 
    = -  \int \rho_{D^\star}(\rr) \, \rr \cdot 
    \nabla v_{\mathrm{L,xc}}^\star(\rr) \rmd \rr,
\end{equation}
which establishes the desired relation between the xc potential and the energy functionals.

To compare this result with the standard density-functional version, 
we define
\begin{equation}\label{eq:ExcOEP}
    E_{\mathrm{xc},\mu}^\mathrm{OEP}(D) = 
     \Gxc(D) + (1+\mu)T_\mathrm{c}(D),
\end{equation}
where we recall that $ \Gxc(D)$ does not contain any contribution from kinetic energy. Thus, Eq.~\eqref{eq:VR-loc-xc} above can be rewritten as
\begin{equation}
    E_{\mathrm{xc},\mu}^\mathrm{OEP}(D^\star) + (1+\mu) T_\mathrm{c}(D^\star)
    = -  \int \rho_{D^\star}(\rr) \, \rr \cdot 
    \nabla v_{\mathrm{L,xc}}^\star(\rr) \rmd \rr.
\end{equation}
This virial expression is in analogy with the standard DFT version \cite[Eq.~(3)]{LevyPerdewPhysRevA1985}, except that presence of the factor $(1+\mu)$ in front of $T_\mathrm{c}$. Since the relation was derived under the assumption that the non-local functional $G_{\mathrm{Hxc}}(D)$ is exact, deviations from this relation provides one possible measure of the error made by an approximate functional.

\vspace{.5cm}
\section{Conclusions}

By adding the excess energy of an auxiliary noninteracting system to the standard energy expression, we obtain a particularly simple and transparent formulation. It leads to a joint minimization principle for $(D,\vL)$, where the minimization over $\vL$ is always a convex optimization problem. The fact that it is a joint optimization problem leads to fully self-consistent calculations, where both the non-local and local potentials are updated.

A technical advantage is that there is no need to refer to the Hohenberg--Kohn mapping and any failures of noninteracting $v$-representability therefore have relatively benign consequences, leading only to an energy gap between $E^1(\vext)$ and the local approximation $\Eloc^1(\vext)$.
The degree of noninteracting $v$-representability of the ground state of the $\mu=0$ orbital-dependent model can be quantified by the gap $\Eloc^1(\vext) - E^1(\vext)$. Moreover, the density $\rho_{D^{\star}}$ obtained at $\mu > 0$ is guaranteed to have a noninteracting $v$-representable projection.

No approximations beyond basis set expansion and possible failures of $v$-representability are involved. In particular, no approximation akin to the Krieger--Li--Iafrate approximation~\cite{KRIEGER_PRA46_5453} is required. Moreover, the formulation is very general in that it is not restricted to exact exchange, but applies to arbitrary orbital-dependent Hxc functionals. It is also general in that extensions to other density-functional frameworks, such as collinear spin-DFT, noncollinear spin-DFT~\cite{KUBLER_JPF_18_469}, and current-density functional theory~\cite{VIGNALE_PRL59_2360,VIGNALE_PRB37_10685}, are straightforward.

\section*{Acknowledgments}
This work was supported by the Norwegian Research Council through the
CoE Hylleraas Centre for Quantum Molecular Sciences Grant No.\ 262695
and through Grant No.~240674.

\end{document}